\documentclass[prb,aps,amsmath,amssymb,showpacs,floatfix,twocolumn]{revtex4}
\usepackage{graphicx,epsfig}
\usepackage[usenames]{color}

\newcommand{\be}{\begin{equation}}
\newcommand{\ee}{\end{equation}}
\newcommand{\bea}{\begin{eqnarray}}
\newcommand{\eea}{\end{eqnarray}}
\newcommand{\ul}{\underline}

\begin{document}
\title{FLEX-description of the spectral functions near singlet-triplet transition} 
\author{B. Horv\'ath}
\affiliation{
Theoretical Physics Department, Institute of Physics, Budapest University of 
Technology and Economics, Budafoki \'ut 8, H-1521 Hungary}

\begin{abstract}
In a previous article\cite{us}, we have investigated the non-equilibrium two-level Anderson model with a simple iterative perturbation theory. 
Here we use here a more sophisticated perturbative method, the fluctuation-exchange (FLEX) approximation. The great advantage of FLEX is its 
\textit{conserving} nature, and that it can describe well the Kondo energy scale, the Kondo-temperature, $T_{K}$. As it was expected from the 
results obtained with iterative perturbation theory, the FLEX description can give back also the relevant features of the spectral properties. 
\end{abstract}

\maketitle

\section{Introduction}

In the past decades the investigation of electronic transport through quantum nanostructures gained a lot of interest both experimentally and 
theoretically.\cite{rev1,rev2,rev3,rev4,rev5,rev6} 
By the development of nanotechnology the properties of these structures gets more controllable. 
In most of the cases, the nanostructures are investigated under out of equilibrium conditions in the experiments.  
In the case of a quantum dot, a single atom or molecule between the biased 
leads, the system can be described well with the non-equilibrium Anderson model.
\cite{noneqand0,noneqand1,noneqand2,noneqand3,noneqand4,noneqand5,noneqand6,noneqand7,noneqand8,aligia} 
The dynamical features of the quantum dot show an odd-even asymmetric behavior with the number of electron on the dot. 
If there are odd number of electrons residing on the dot, spin$-\frac{1}{2}$ Kondo effect can be observed. 
By even occupation, on which we will focus in the present paper, the situation is more complex, the triplet Kondo effect\cite{Wiel,Kogan,granger} and the 
Kondo-effect associated with singlet-triplet degeneracy
\cite{sasaki} 
were found in vertical\cite{sasaki} and lateral\cite{Wiel,granger,zarand} quantum dots (QDs).
The lateral and vertical QDs have different geometry, 
with different number of modes connecting the dot levels to the leads, while in the vertical case the number of modes is more than one, 
in the lateral case one or two modes connects the QD to the conduction electrons of the leads. 
At the triplet side of singlet-triplet transition, we can speak about 
fully screened\cite{Wiel, zarand} (2 modes) and underscreened spin$-1$ Kondo-effect\cite{roch,noneqand6} (1 mode), 
the transport properties are totally different in this two case. 

To the description of the quantum dots under out of equilibrium conditions, a great number of 
perturbative and non-perturbative theoretical scheme of Anderson model was generalized from equilibrium. 
The most effective perturbative treatments are 
the iterative perturbative approaches which was developed also for single\cite{aligia,noneqand8} and multi-level\cite{us,yeyati2} Anderson models. 
These descriptions are stable even at rather high interaction parameters and have the great advantage that they can be easily generalized to 
non-equilibrium using Keldysh formalism,\cite{rammer} but cannot give back precisely the 
width of the Kondo-peak. (A detailed description of the adventages and disadventages of this method is given in Ref.\onlinecite{us}.)

There are a lot of non-perturbative treatments also, which were generalized to out of equilibrium from successful equilibrium treatments. 
The Bethe Ansatz approach was applied to finite biases,\cite{andrei} 
and recently it was worked out for the Anderson model.\cite{noneqand5} 
The Numerical Renormalization Group technique was 
generalized with applying a single-particle scattering basis.\cite{anders} Other Renormalization Group techniques also had successes in 
the description of a steady-state non-equilibrium state.\cite{egger} Also numerical methods such as quantum Monte Carlo\cite{egger} 
and Density Matrix RG\cite{Scholl} was generalized to out of equilibrium 
and was applied to interacting nanostructure transport problems, but these schemes suffer from the difficulties inherited from equilibrium. 
The powerful technique of the Dynamical Mean Field theory was extended to non-equilibrium.\cite{freericks} 
Also a yet another technique, the iterative summation of real-time path integrals was developed\cite{egger}, 
and its validity has been confirmed by a detailed comparison to approximative approaches.

The relative simplicity of generalization to non-equilibrium of the perturbative schemes is also present in fluctuation-exchange (FLEX) 
approximation,\cite{bickers} but FLEX is a \textit{conserving}\cite{kadanoff,baym,bickers} approximation for the whole parameter space, 
and describes well the width of the Kondo peak in the equilibrium spectral function. It cannot describe the charge-excitations (the so-called 
Hubbard-peaks) in the spectral function\cite{white}, but we assume that in the non-equilibrium properties this does not play an important role. 
FLEX sums up 
some classes of diagrams to infinite order, 
we chose to sum up the particle-hole type diagrams, in the present case of the multi-level Anderson model. 
The other summable diagram types do not play a significant role in this problem. 
We describe here lateral and vertical two-level quantum dots with approximately two electrons in average on them, while these systems show many features 
of the singlet-triplet transition in and out of equilibrium.\cite{zarand,Pust1,Pust2,Wiel,granger}

The present paper is a second part of a two part series, in the first part\cite{us} (called paper I in the following) 
we have investigated the same system with the scheme of the 
iterative perturbation theory (IPT).\cite{kotliar,aligia,noneqand8} As we found there, that IPT can capture many properties well in equilibrium, and it gives satisfactory 
results even in the non-equilibrium limit. Based on the successes of that perturbative description, we assume that also the FLEX description 
can give good equilibrium and non-equilibrium results. 

The paper is organized as follows. In Sec.~\ref{model} we introduce the model of two level non-equilibrium Anderson model, 
and the vertex perturbation theory done in the Coulomb interaction and the Hund's rule coupling. In Sec.~\ref{flexsec} 
we show how to apply the fluctuation-exchange approximation to this model. In Sec.~\ref{iter} we show the details of the FLEX iteration loop, and in 
Sec.~\ref{limits} we write about the limitations of this scheme. In Sec.~\ref{results} we detail the results obtained for the spectral functions, 
in Sec.~\ref{sym} for the symmetrically hybridized case, while in Sec.~\ref{assym} for the asymmetrically hybridized case.  We conclude in 
Sec.~\ref{conclu} and give some calculation details in Appendix~\ref{hybrg} and \ref{selfapp}.

\section{Theoretical framework}

\subsection{Model}\label{model}

We describe an out of equilibrium quantum dot with the two-level Anderson model 
as we have done it in paper I. 
The Hamiltonian of the system can be splitted to a non-interacting and interacting term: $H = H_{0}+H_{\rm int}$, where the non-interacting term, 
$H_{0}$ can be further divided onto three terms similarly to the Eq. (1) of paper I,
\be
 H_{0}=H_{\rm cond}+H_{\rm hyb}+H_{0,\rm dot}\label{ham}. 
\ee 
In this expression $H_{\rm cond}$ describes the conduction electrons of the leads, $H_{\rm hyb}$ accounts for the tunneling between the conduction 
electrons and the electrons residing on the dot. This two part is different in the case of lateral and vertical quantum dot. 

For lateral QD, one electron mode is coupling to the dot per lead because the lateral QD is close to pinch-off, 
the Hamiltonians are (cf. Eqs.~(3)-(4) of paper I), 
\begin{eqnarray}
   H^{\rm lat}_{\rm cond} & = & \sum_{\xi,\alpha,\sigma}
\xi_{\alpha}c_{\xi\alpha\sigma}^{\dagger}c_{\xi\alpha\sigma}\label{latcond}\;,\\
   H^{\rm lat}_{\rm hyb} & = & \sum_{\alpha,i,\xi,\sigma}t_{\alpha
     i}(c_{\xi\alpha\sigma}^{\dagger}d_{i\sigma}+h.c.)\;.
\end{eqnarray}
$c_{\xi\alpha\sigma}^{\dagger}$ is the creation operator of a conduction electron in
the left or right lead  ($\alpha\in(L,R)$) with energy  
$\xi_{\alpha}=\xi+\mu_{\alpha}$, ($\mu_{\alpha}=eV_{\alpha}$ is the bias applied on lead $\alpha$, the non-equilibrium condition is taken 
into account in a different shift of the chemical potential in each lead) 
and spin $\sigma$. $d_{i\sigma}$ destroys an electron with spin $\sigma$ on the dot 
level  $i\in(+,-)$ and $t_{\alpha i}$ is the tunneling matrix element connecting the lead $\alpha$ and dot level $i$,  
$\ul{\ul{t}}$ is assumed to be spin- and energy independent. 

In the vertical case, the dot is connected to the leads with a relatively large surface, therefore there are many different conduction channels 
coupling to a given dot state. For each dot state $i$ and lead $\alpha$ one can construct a simple linear combination of the modes, 
$c_{\xi,i\alpha\sigma}^{\dagger}$, that hybridizes with $d_{i\sigma}$, and assumed to be independent for $i=\pm$. The Hamiltonians are 
(cf. Eq. (5)-(6) of paper I), 
\begin{eqnarray}
   H^{\rm vert}_{\rm cond} & = & \sum_{\xi,i,\alpha,\sigma}\xi_{\alpha}c_{\xi i\alpha\sigma}^{\dagger}c_{\xi i\alpha\sigma}\label{vertcond}\;,\\
   H^{\rm vert}_{\rm hyb} & = & \sum_{\alpha,i,\xi,\sigma}
t_{\alpha i}(c_{\xi i\alpha\sigma}^{\dagger}d_{i\sigma}+h.c.)\;.
\end{eqnarray}
Eqs. (\ref{latcond}) and (\ref{vertcond}) shows that the interaction between the conduction electrons is neglected. 

The last term of Eq. (\ref{ham}) describe the bare dot levels,  
\begin{equation}
H_{0,\rm dot} = \sum_{ij,\sigma}\varepsilon_{ij}d_{i\sigma}^{\dagger}d_{j\sigma}\;,
\end{equation}
The eigenvalues of the $\ul{\ul{\varepsilon}}$ are the energies 
corresponding to $i=\pm$. 

Throughout this paper we consider a special completely symmetric tunneling matrix, similarly to the IPT description of paper I, 
because this case captures most of the spectral and transport properties obtained in experiments for a two-level QD. 
We assume, that one of the dot levels is even ($+$), the other is 
odd ($-$) under reflection. Under these conditions, the tunneling matrix has simple structure: $t_{L,+}=t_{R,+}$ and 
$t_{L,-}=-t_{R,-}$.\cite{zarand,izumida} In one of the cases we specialize further this situation with $|t_{\alpha,+}|=|t_{\alpha,-}|$ 
which gives a particle-hole symmetric case with exactly two electrons on the dot. 
By these values of the parameters, we define here the hybridization parameter, the $\ul{\ul{\Gamma}}$-matrix, connected 
to the level width of the dot states, 
\begin{equation}
 \Gamma_{ij}=2\pi\rho(0)\sum_{\alpha}t_{\alpha i}t_{\alpha j}^{*}\label{gammadef}\;, 
\end{equation}
where $\rho(0)$ is the spectral density of the conduction electrons at Fermi-energy, we take it $\rho(0)=1$. 
Here is given a general form, but we have to note that by the above detailed special parameter values, only the diagonal elements are finite of the matrix.
The diagonal elements will be called, for the sake of brevity, $\Gamma_{+}\equiv\Gamma_{++}$ and $\Gamma_{-}\equiv\Gamma_{--}$. 

Using the expression (7) of paper I, the electron-electron interaction can described with
\begin{equation}
 H_{\rm int} =
 \frac{U}{2}\left(\sum_{i\sigma}n_{i\sigma}-2\right)^{2}-J\;\vec{S}^{2}\;,
\label{intham}
\end{equation}
where the occupation number is $n_{i\sigma}=d_{i\sigma}^{\dagger}d_{i\sigma}$ and the spin is
$\vec{S}=\frac{1}{2}\sum\limits_{i\sigma\sigma'}d_{i\sigma}^{\dagger}
\vec{\sigma}_{\sigma\sigma'}d_{i\sigma'}$ with the Pauli matrices,  $\vec{\sigma}=(\sigma_x,\sigma_y,\sigma_z)$.
$U$ is the on-site Coulomb-interaction, while $J>0$ denotes the ferromagnetic Hund's rule coupling. 
Eq.~(\ref{intham}) contain explicitly that the relevant case of this description is that, when there is (approximately) two electrons on the dot. 

To introduce a systematic perturbation theory, which treats the on-site Coulomb interaction and Hund's rule coupling on equal 
footing, we write Eq.~(\ref{intham}) to normal ordered form, 
\begin{equation}
 H_{\rm int}= :H_{\rm int}: -\left(\frac {3U} 2 +    
\frac {3J} 4\right)  \sum_{i\sigma}n_{i\sigma} \;, 
\label{:intham:}
\end{equation}
where the normal order part will play the role of a perturbation, the second term is a simple renormalization of the single electron levels. 

\begin{figure}
\includegraphics[width=240pt,clip=true]{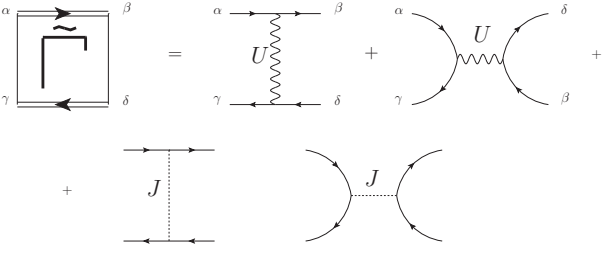}
\caption{The construction of the particle-hole like vertex used by building FLEX self-energies}\label{phvertex}
\end{figure}

The normal ordered term of Eq.~(\ref{:intham:}) can be rewritten using an antisymmetrized interaction vertex of particle-hole type, 
$\tilde{\Gamma}_{i\sigma\;m\tilde\sigma'}^{j\sigma'\;n\tilde\sigma}$ shown in Fig.~\ref{phvertex}, 
as 
\begin{equation}
 \tilde{H}_{\rm int}=\sum_{ijmn,\sigma\sigma'\tilde\sigma \tilde\sigma'}
\frac{1}{4}\;\tilde\Gamma_{i\sigma\;m\tilde\sigma'}^{j\sigma'\;n\tilde\sigma}
d_{j\sigma'}^{\dagger}d_{m\tilde\sigma'}^{\dagger}d_{n\tilde\sigma}d_{i\sigma}\;, 
\label{eq:vert_ham}
\end{equation}
which is very similar to the expression (11) of paper I, but in opposition to that, this expression uses particle-hole vertices, 
instead of particle-particle vertices. $\tilde{\Gamma}_{i\sigma\;m\tilde\sigma'}^{j\sigma'\;n\tilde\sigma}$ is antisymmetrical under 
exchanges $i\sigma\leftrightarrow m\tilde\sigma'$ and $j\sigma'\leftrightarrow n\tilde\sigma$, the finite matrix 
elements of the interaction vertex can be easily found. 
Creating the diagrams with this vertex function, the details of the on-site Coulomb interaction and the Hund's rule coupling will be hidden in 
the vertex (it can be resolved with the substitution of the diagrammatic form of the vertex in Fig.~\ref{phvertex}), and the equal-footing treatment of 
$U$ and $J$ is ensured. 

\subsection{Out of equilibrium fluctuation exchange approximation}\label{flexsec}

In this section we will show how to apply the fluctuation-exchange approximation (FLEX) to the description of a two-level QD under 
non-equilibrium conditions. It is shown how to calculate the out of equilibrium Green's functions ($G^{\kappa,\kappa'}$, 
$\kappa$ and $\kappa'$ sign the branches of the Keldysh-contour) 
with FLEX-approximation in the self energy. 

The great advantage of FLEX that it sums up a group of diagrams to the infinite order. In the case of the two-level Anderson model 
the so-called particle-hole diagrams are relevant, therefore, we consider only this group of diagrams in the following. 
We can build equivalent diagrams with the two kind of interaction vertices which was used in paper I and that in the 
recent work in Fig.~\ref{phvertex}. 
In opposition with the particle-particle type vertex of paper I, the particle-hole type behaves
 $\tilde{\Gamma}_{\alpha\gamma}^{\beta\delta}=-\tilde{\Gamma}_{\alpha\beta}^{\gamma\delta}$ 
by exchange of the two outgoing legs. Where $\alpha, \beta, \gamma, \delta$ are compound indices, 
$\alpha=(l,\sigma,\kappa)$. The vertex has simple behavior in Keldysh-space:
\be
\tilde{\Gamma}_{l_1,\sigma_1,\kappa_1\;l_2,\sigma_2,\kappa_2}^{l_3,\sigma_3,\kappa_3\;l_4,\sigma_4,\kappa_4}
=S(\kappa_1)\delta_{\kappa_1\kappa_2\kappa_3\kappa_4}\tilde{\Gamma}_{l_1,\sigma_1\;l_2,\sigma_2}^{l_3,\sigma_3\;l_4,\sigma_4}\;.
\ee

While the interaction vertices are spin-conserving, only vertex elements with special spin arrangement are finite. 
We arrange the vertex elements in a matrix based on the spin arrangements. The rows (upper pair of spins in the vertex) and the columns 
(lower pair of spins in the vertex) have the spins $\uparrow\uparrow$, $\uparrow\downarrow$, $\downarrow\uparrow$ and $\downarrow\downarrow$ 
respectively. We emphasize that in each position of this spin matrix, a matrix of the level and Keldysh indices resides. 
Because of the special spin arrangements this matrix can be described by two independent matrix variables as follows,
\be
\ul{\ul{\tilde{\Gamma}}}_{\sigma_1\sigma_2}^{\sigma_3\sigma_4}=
\begin{bmatrix} \frac{1}{2}\left(\ul{\ul{\tilde{\Gamma}}}_{s}+\ul{\ul{\tilde{\Gamma}}}_{t}\right) 
 & 0 & 0 & \frac{1}{2}\left(\ul{\ul{\tilde{\Gamma}}}_{s}-\ul{\ul{\tilde{\Gamma}}}_{t}\right)\\
0 & \ul{\ul{\tilde{\Gamma}}}_{t} & 0 & 0 \\ 0 & 0 & \ul{\ul{\tilde{\Gamma}}}_{t} & 0 \\
\frac{1}{2}\left(\ul{\ul{\tilde{\Gamma}}}_{s}-\ul{\ul{\tilde{\Gamma}}}_{t}\right) 
 & 0 & 0 & \frac{1}{2}\left(\ul{\ul{\tilde{\Gamma}}}_{s}+\ul{\ul{\tilde{\Gamma}}}_{t}\right)\;
\end{bmatrix}
.
\ee
If we substitute in the finite vertex elements, we can find the finite elements of these singlet and triplet matrices occurring in the matrix
\bea
\ul{\ul{\tilde{\Gamma}}}_{t} & = & \ul{\ul{\tilde{\Gamma}}}_{\uparrow\uparrow}^{\uparrow\uparrow}-
\ul{\ul{\tilde{\Gamma}}}_{\downarrow\downarrow}^{\uparrow\uparrow}\;,\\
\ul{\ul{\tilde{\Gamma}}}_{s} & = & \ul{\ul{\tilde{\Gamma}}}_{\uparrow\uparrow}^{\uparrow\uparrow}+
\ul{\ul{\tilde{\Gamma}}}_{\downarrow\downarrow}^{\uparrow\uparrow}\;. 
\eea
With this classification of the spin, the whole treatment simplifies. Calculating with this singlet-triplet matrices we have to treat only Keldysh 
and level indices in the calculations. 

\begin{figure}
\centering\includegraphics[width=100pt,clip=true]{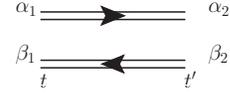}
\caption{Diagrammatic representation of $\Pi$-function}\label{pi}
\end{figure}

\begin{figure}
\centering\includegraphics[width=240pt,clip=true]{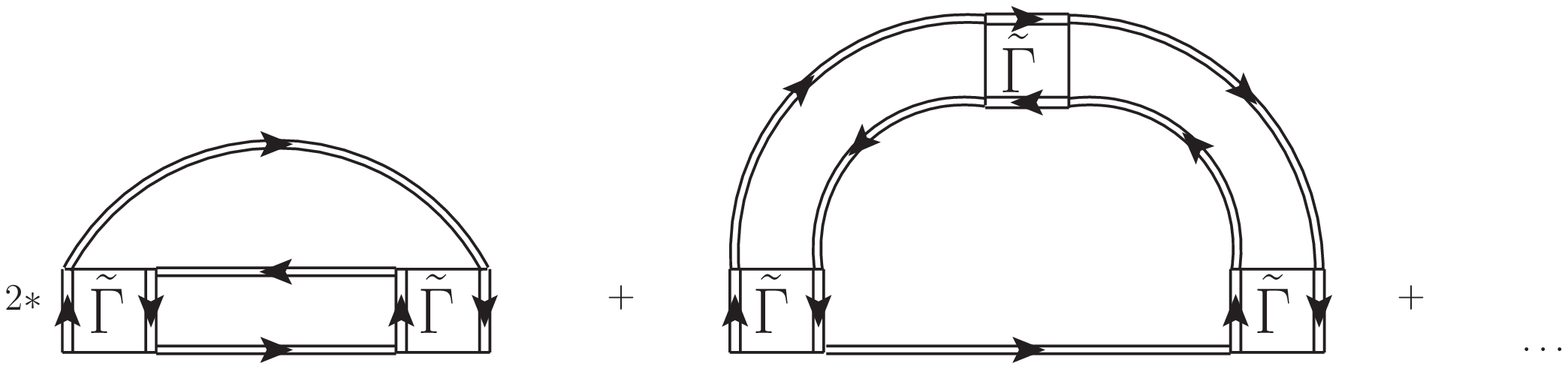}
\caption{FLEX self-energy}\label{selfen}
\end{figure}

\begin{figure}
\centering\includegraphics[width=240pt,clip=true]{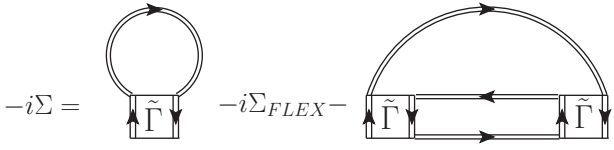}
\caption{Full self-energy}\label{fullself}
\end{figure}

By the construction of the diagrams, it is convenient to introduce a new quantity, the so-called polarization function, defined diagrammatically 
in Fig.~\ref{pi},  
\be
 {\Pi^{(0)}}_{\alpha_1\beta_1}^{\alpha_2\beta_2}(t-t')=i^{2} G_{\alpha_1}^{\alpha_2}(t-t')G_{\beta_2}^{\beta_1}(t'-t)\;,\label{polariz}
\ee
where $G_{\alpha_1}^{\alpha_2}$ is the full Green's function.
Using the polarization function and the interaction vertex, it is very straightforward to construct the full FLEX self-energy, 
shown in Fig.~\ref{selfen}. 
In the figure the hierarchical order of the self-energy can be observed. It is relevant 
to note that using these definitions, the second order diagram is over-counted in the FLEX self-energy, this will have to be compensated in the 
full self-energy, as shown in the diagrammatic form of it in Fig.~\ref{fullself} (all the other orders  of the FLEX self-energy are not over-counted).

\begin{figure}
\centering\includegraphics[width=240pt,clip=true]{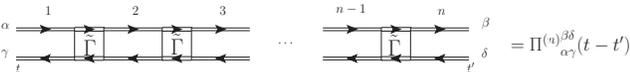}
\caption{$n$th order polarization function, $\Pi^{(n)}$}\label{pin}
\end{figure}

The $n+1$-th order diagram of the FLEX self-energy can be described in the time-domain
\bea
& &{\Sigma_{\alpha}^{\beta}}_{FLEX}^{(n+1)}(t-t')=(-i)^{n+1}\idotsint\limits_{-\infty}^{\infty}dt_{1}\dots dt_{n}\nonumber\\
& &\tilde{\Gamma}_{\alpha\tilde{\alpha}}^{\tilde{\alpha}_{0}\tilde{\beta}_{0}}
{\Pi^{(0)}}_{\tilde{\alpha}_{0}\tilde{\beta}_{0}}^{\alpha_{1}\beta_1}(t-t_1)
\tilde{\Gamma}_{\alpha_1\beta_1}^{\tilde{\alpha}_{1}\tilde{\beta}_{1}}
{\Pi^{(0)}}_{\tilde{\alpha}_{1}\tilde{\beta}_{1}}^{\alpha_{2}\beta_2}(t_1-t_2)\times\dots\times\nonumber\\ 
&\times& {\Pi^{(0)}}_{\tilde{\alpha}_{n-1}\tilde{\beta}_{n-1}}^{\alpha_{n}\beta_n}(t_n-t')
\tilde{\Gamma}_{\alpha_n\beta_n}^{\beta\tilde{\beta}}G_{\tilde{\alpha}}^{\tilde{\beta}}(t-t')\;,
\eea
where the integration goes over the inner variables, $t_{1}\dots t_{n}$. The part of this expression which depends on the inner variables can be 
organized in a new quantity 
\bea
& & {\Pi^{(n)}}_{\tilde{\alpha}_{0}\tilde{\beta}_{0}}^{\alpha_{n}\beta_n} = (-i)^{n-1}
\idotsint\limits_{-\infty}^{\infty}dt_{1}\dots dt_{n}\times\nonumber\\
& \times & {\Pi^{(0)}}_{\tilde{\alpha}_{0}\tilde{\beta}_{0}}^{\alpha_{1}\beta_1}(t-t_1)
\tilde{\Gamma}_{\alpha_1\beta_1}^{\tilde{\alpha}_{1}\tilde{\beta}_{1}}
{\Pi^{(0)}}_{\tilde{\alpha}_{1}\tilde{\beta}_{1}}^{\alpha_{2}\beta_2}(t_1-t_2)\times\dots\nonumber\\
& \times & {\Pi^{(0)}}_{\tilde{\alpha}_{n-1}\tilde{\beta}_{n-1}}^{\alpha_{n}\beta_n}(t_n-t')\;,
\eea
diagrammatically shown in Fig.~\ref{pin}.
Using this expression, we can write down $n+1$th order self-energy in a simpler form, 
\bea
& & {\Sigma_{\alpha}^{\beta}}_{FLEX}^{(n+1)}(t-t')=\nonumber\\
&=&-\tilde{\Gamma}_{\alpha\tilde{\alpha}}^{\tilde{\alpha}_{0}\tilde{\beta}_{0}}
{\Pi^{(n)}}_{\tilde{\alpha}_{0}\tilde{\beta}_{0}}^{\alpha_{n}\beta_{n}}(t-t')
\tilde{\Gamma}_{\alpha_{n}\beta_{n}}^{\beta\tilde{\beta}}G_{\tilde{\alpha}}^{\tilde{\beta}}(t-t')\;.
\eea
The polarization functions have the same structure in spin as the interaction vertices, therefore we can use the same matrix 
description for them, 
\be
\ul{\ul{\Pi}}_{\sigma_1\sigma_2}^{\sigma_3\sigma_4}=
\begin{bmatrix} \frac{1}{2}\left(\ul{\ul{\Pi}}_{s}+\ul{\ul{\Pi}}_{t}\right) 
 & 0 & 0 & \frac{1}{2}\left(\ul{\ul{\Pi}}_{s}-\ul{\ul{\Pi}}_{t}\right)\\
0 & \ul{\ul{\Pi}}_{t} & 0 & 0 \\ 0 & 0 & \ul{\ul{\Pi}}_{t} & 0 \\
\frac{1}{2}\left(\ul{\ul{\Pi}}_{s}-\ul{\ul{\Pi}}_{t}\right) 
 & 0 & 0 & \frac{1}{2}\left(\ul{\ul{\Pi}}_{s}+\ul{\ul{\Pi}}_{t}\right)
\end{bmatrix}
\;.
\ee
Considering only the different inner part described by $\ul{\ul{\Pi}}^{(n)}$ in each order, it is transparent that it can be summed up easily 
to the infinite order, because they give a geometrical order:
\bea
\ul{\ul{\Pi}}(\omega) & = & \sum_{n=0}^{\infty}\ul{\ul{\Pi}}^{(n)} = 
\ul{\ul{\Pi}}^{(0)}(\omega) + \ul{\ul{\Pi}}^{(0)}(\omega)(-i\ul{\ul{\tilde{\Gamma}}})\ul{\ul{\Pi}}^{(0)}(\omega) + 
\dots\nonumber\\
& = & \ul{\ul{\Pi}}^{(0)}(\omega)\left[\ul{\ul{1}}+i\ul{\ul{\tilde{\Gamma}}}\ul{\ul{\Pi}}^{(0)}(\omega)\right]^{-1}\;.\label{pifull}
\eea
Using the singlet-triplet description in the polarization matrix ($\ul{\ul{\Pi}}_{s}$, $\ul{\ul{\Pi}}_{t}$), similarly to the interaction 
vertices, Eq.~(\ref{pifull}) reduces to the two equations as follows, 
\bea
 \ul{\ul{\Pi}}_{s}(\omega) 
  & = & \ul{\ul{\Pi}}_{s}^{(0)}(\omega)\left[\ul{\ul{1}}+i\ul{\ul{\tilde{\Gamma}}}_{s}\ul{\ul{\Pi}}_{s}^{(0)}(\omega)\right]^{-1}\;,\label{sing}\\
 \ul{\ul{\Pi}}_{t}(\omega)
  & = & \ul{\ul{\Pi}}_{t}^{(0)}(\omega)\left[\ul{\ul{1}}+i\ul{\ul{\tilde{\Gamma}}}_{t}\ul{\ul{\Pi}}_{t}^{(0)}(\omega)\right]^{-1}\;.\label{trip}
\eea

The inner structure of $\ul{\ul{\Pi}}_{s,t}$ and $\ul{\ul{\tilde{\Gamma}}}_{s,t}$ matrices can be described by a compound index, 
which contain level and Keldysh indices. If we describe $\Pi_{i}^{j}$ as a $2$-index matrix, the compound indices are 
\bea
 i & = & 1:\;\left(++\;1\right);\quad i=2:\;\left(+-\;1\right);\quad\nonumber\\ 
 i & = & 3:\;\left(-+\;1\right);\quad i=4:\;\left(--\;1\right);\quad\nonumber\\
 i & = & 5:\;\left(++\;2\right);\quad i=6:\;\left(+-\;2\right);\quad\nonumber\\ 
 i & = & 7:\;\left(-+\;2\right);\quad i=8:\;\left(--\;2\right),\quad\nonumber
\eea
where $l=\pm$ are the level indices, $\kappa=1,2$ are the Keldysh indices. 

While the inner part brings the order of the diagram, we can obtain the full FLEX self-energy by summing up these expressions to infinite order, 
\bea
 \Sigma_{\rm FLEX} & = & \sum_{n=1}^{\infty}\Sigma_{\rm FLEX}^{(n+1)}\;,\\
 {\Sigma_{\alpha}^{\beta}}_{\rm FLEX}(t-t') & = & -\tilde{\Gamma}_{\alpha\tilde{\alpha}}^{\tilde{\alpha}_{0}\tilde{\beta}_{0}}
\Pi_{\tilde{\alpha}_{0}\tilde{\beta}_{0}}^{\tilde{\alpha}_{1}\tilde{\beta}_{1}}(t-t')\times\\
&\times&\tilde{\Gamma}_{\tilde{\alpha}_{1}\tilde{\beta}_{1}}^{\beta\tilde{\beta}}G_{\tilde{\alpha}}^{\tilde{\beta}}(t-t')\;.
\eea

\subsection{Details of the FLEX iteration}\label{iter}

In a numerical calculation the Green's functions are represented on a 
finite, discrete frequency mesh characterized by two basic quantities, 
the number of the grid points ($N$) and the width of the frequency range 
the function was represented on ($\Omega$). 
We used an equidistant mesh with distance between the grid points
$\Delta\omega=\Omega/N$.
In the present calculations we use $\Omega=1000$ and $N=2^{17}$ to obtain spectral functions. 

Here we write down in detail one iteration loop of the FLEX approximation. 
It is relevant to emphasize in the beginning of this section, that the full Green's function is iterated in the FLEX scheme, every points of the function 
have to converge. We use the $\omega$-space Green's functions to check the convergence. 
As a convergence criterion, we are searching for the largest difference of the previous and the new Green's functions as a function of 
$\omega$-points.  
The initialization of the full Green's function at the 
beginning of each step
\bea
 {G^{0}}_{\alpha_1}^{\alpha_2}(\omega) & = & g_{\alpha_1}^{\alpha_2}(\omega)\;,\\
 {G^{n\neq0}}_{\alpha_1}^{\alpha_2}(\omega) & = & {G^{n-1}}_{\alpha_1}^{\alpha_2}(\omega)\;,
\eea
where the number of the iteration signed in the upper index. 
In the $0$th order we initialize the Green's function with the hybridized Green's function, the calculation of it is introduced in the 
Appendix \ref{hybrg}. To define the polarization we need Green's functions in '$t$'-space, using Fourier transformation:
\be
 {G^{n}}_{\alpha_1}^{\alpha_2}(t) = IFFT({G^{n}}_{\alpha_1}^{\alpha_2}(\omega))\;.
\ee
Using a flip, we calculate ${G^{n}}_{\alpha_1}^{\alpha_2}(t)\rightarrow{G^{n}}_{\alpha_1}^{\alpha_2}(-t)$, and the polarization in time-space is obtained 
from Eq.~(\ref{polariz}). 
In the following, we will use matrix notation for the polarization and we will need a singlet and triplet part of $\ul{\ul{\Pi}}^{(0)}$, 
which will have only level and Keldysh indices. The Green's functions are diagonal in spin, therefore by definition, the $0$th order 
polarization has the same value in the singlet and triplet channel as follows, 
\bea
 \ul{\ul{\Pi}}_{s}^{(0)} & = & \ul{\ul{\Pi}}_{\uparrow\uparrow}^{\uparrow\uparrow}+\ul{\ul{\Pi}}_{\uparrow\uparrow}^{\downarrow\downarrow}= 
\ul{\ul{\Pi}}_{\uparrow\uparrow}^{\uparrow\uparrow}\;,\\
 \ul{\ul{\Pi}}_{t}^{(0)} & = & \ul{\ul{\Pi}}_{\uparrow\downarrow}^{\uparrow\downarrow}\;,\\
 \ul{\ul{\Pi}}_{s}^{(0)} & = & \ul{\ul{\Pi}}_{t}^{(0)}\;.
\eea
To obtain the interacting polarization function we need $\ul{\ul{\Pi}}_{s,t}^{(0)}(\omega)=FFT(\ul{\ul{\Pi}}_{s,t}^{(0)}(t))$, 
and using the above defined $\ul{\ul{\tilde{\Gamma}}}_{s}$ and $\ul{\ul{\tilde{\Gamma}}}_{t}$, we can create the fully interacting 
$\ul{\ul{\Pi}}$ following Eqs. (\ref{sing}) and (\ref{trip}). 

To construct the components of the self-energy the polarization function is needed in time-space, 
\be
 \ul{\ul{\Pi}}_{s,t}(t)=IFFT(\ul{\ul{\Pi}}_{s,t}(\omega))\;, 
\ee
using this function, the full self-energy can be constructed from four parts
\bea
 {\Sigma^{n}}_{\alpha}^{\beta}(t) & = & {{\Sigma_{FLEX}}^{n}}_{\alpha}^{\beta}(t)+{{\Sigma_{ct}}^{n}}_{\alpha}^{\beta}(t)+\nonumber\\
& + & {{\Sigma^{(1)}}^{n}}_{\alpha}^{\beta}(t)-{{\Sigma^{(2)}}^{n}}_{\alpha}^{\beta}(t)\;,
\eea
where the components are
\bea
 {\Sigma_{FLEX}}_{\alpha}^{\beta}(t) & = & -\sum_{\tilde{\alpha}\tilde{\beta}}
\left(\ul{\ul{\tilde{\Gamma}}}\ul{\ul{\Pi}}(t)\ul{\ul{\tilde{\Gamma}}}\right)_{\alpha\tilde{\alpha}}^{\beta\tilde{\beta}}
G_{\tilde{\alpha}}^{\tilde{\beta}}(t)\;,\label{flse}\\
{{\Sigma_{lin}}^{n}}_{\alpha}^{\beta}(t) & = & -\left(\frac{3U}{2}+\frac{3J}{4}\right)S(\kappa)\delta_{\alpha\beta}\delta(t)\;,\label{linpart}\\
{{\Sigma^{(1)}}^{n}}_{\alpha}^{\beta}(t) & = & \delta_{\kappa\kappa'}\sum_{\tilde{\alpha}\tilde{\beta}}
\ul{\ul{\tilde{\Gamma}}}_{\alpha\beta}^{\tilde{\alpha}\tilde{\beta}}
{n}_{\tilde{\alpha}}^{\tilde{\beta}}\label{1stord}\;,\\
{{\Sigma^{(2)}}^{n}}_{\alpha}^{\beta}(t) & = & -\sum_{\tilde{\alpha}\tilde{\beta}}
\left(\ul{\ul{\tilde{\Gamma}}}\ul{\ul{\Pi}}^{(0)}(t)\ul{\ul{\tilde{\Gamma}}}\right)_{\alpha\tilde{\alpha}}^{\beta\tilde{\beta}}
G_{\tilde{\alpha}}^{\tilde{\beta}}(t)\label{2se}\;,
\eea
where Eq. (\ref{flse}) describes the FLEX self-energy summed up to infinite order, Eq. (\ref{linpart}) contains the linear shift obtained 
by taking the interaction to normal order, Eq. (\ref{1stord}) describes the first order Hartree diagrams, and Eq. (\ref{2se}) the 
second order diagram, which compensate the over-counting. The FLEX (\ref{flse}), the first order (\ref{1stord}) and the second order 
(\ref{2se}) self-energies can be 
calculated in a simple way using the singlet-triplet representation of the interaction vertex and polarization matrix. 
This procedure is shown in the Appendix~\ref{selfapp}. 

The Dyson equation needs the total self-energy in the '$\omega$'-space, 
\be
 {\Sigma^{n}}_{\alpha}^{\beta}(\omega)=FFT\left({\Sigma^{n}}_{\alpha}^{\beta}(t)\right)\;,
\ee
as follows, 
\be
 {{G^{-1}}^{n+1}}_{\alpha}^{\beta}(\omega) = {g^{-1}}_{\alpha}^{\beta}(\omega) - {\Sigma^{n}}_{\alpha}^{\beta}(\omega)\;,
\ee
where we used the hybridized Green's function ($g$). With an inversion, we can have ${G^{n+1}}_{\alpha}^{\beta}(\omega)$. 
Calculating the convergence parameter in the $n$th loop, 
\be
 \epsilon^{n}=max_{\omega}\left(\sqrt{|{G^{n+1}}_{\alpha}^{\beta}(\omega)-{G^{n}}_{\alpha}^{\beta}(\omega)|^2}\right)\;.
\ee
At the end, we interpolate to have the Green's function for the next iteration, 
\be
 {G^{n+1}}_{\alpha}^{\beta}(\omega)=\alpha {G^{n+1}}_{\alpha}^{\beta}(\omega) + (1-\alpha){G^{n}}_{\alpha}^{\beta}(\omega)\;,
\ee
In the iteration procedure we iterate this loop until the $\epsilon^{n}$ become small enough, in some $n$th order, 
which means that the Green's function is converged. 
When we reach this point we can extract spectral functions from the resulting Green's function from
\begin{equation}
   \rho_{i\sigma}(\omega)=-\frac{1}{\pi}\mathrm{Im}\;G_{\;ii\sigma}^{R}(\omega)\;,
\end{equation}
in the followings in many of the cases, we will use the total spectral function, $\rho_{T}(\omega)=\sum_{i,\sigma}\rho_{i\sigma}(\omega)$, 
or a spectral function only summed up for the spins. 

\subsection{Limitations}\label{limits}

We have seen in paper I that the IPT suffers from instability at large values of  
$U/\Gamma$ and $J/\Gamma$ parameters. Similar kind of instability exists in the FLEX scheme: from arbitrary starting Green's function the 
iteration converges only at small values of $U/\Gamma$ and $J/\Gamma$. The error caused by the not satisfactory behavior of the self-energy contributions 
for large $U/\Gamma$ and $J/\Gamma$ parameter rates, if we use the hybridized Green's function to obtain the self-energy. 
But in opposition to IPT, starting from a converged Green's function, 
high enough interaction parameters can be reached by increasing the interaction parameter in small steps. This means that approximately 
$U/\Gamma\approx2$ can be reach using this procedure starting from a Green's function converged at $U/\Gamma\approx0.75$. 
Also the $J/\Gamma$ rate has to be small enough to have a good convergence for the starting Green's function. 

The unit of energy will be the on-site Coulomb interaction $U$. For the present results of this paper, the calculations have the following two 
part. In the first part, we calculate a FLEX Green's functions starting from the hybridization Green's function corresponding to the parameter 
set with $U=1$ and with comparatively high $\Gamma_{\pm}$ values to reach conductance from this somehow arbitrary Green's function which is 
far from the final FLEX Green's function. After we have the converged FLEX Green's function we can improve the $U/\Gamma_{\pm}$ ratio 
with increasing the value of $U$. We increased $U$ step by step with $\delta U=0.1$ and the final value was $U=2$.

\section{Results and Discussion}\label{results}

When the levels are chosen to be particle-hole symmetric, 
i. e. $\varepsilon_{\pm}=-1.5U\pm\Delta/2$, $\varepsilon_{+-}=\varepsilon_{-+}=0$ and $|t_{\alpha+}|=|t_{\alpha-}|$, two electrons are residing on the dot, 
or when $|t_{\alpha+}|\neq|t_{\alpha-}|$, the number of electrons 
is close to two, and the states can be described well with effectively two electrons.  
These two electrons can form three $S=0$ singlet-state and three $S=1$ triplet state on the dot. 
The singlet states can be represented 
in terms of electron creation operators as 
$|s_{-}\rangle=d^{\dagger}_{-\uparrow}d^{\dagger}_{-\downarrow}|0\rangle$, 
$|s_{+}\rangle=d^{\dagger}_{+\uparrow}d^{\dagger}_{+\downarrow}|0\rangle$,
$|s_{+-}\rangle=\frac{1}{\sqrt{2}}(
d^{\dagger}_{+\uparrow}d^{\dagger}_{-\downarrow}-
d^{\dagger}_{-\uparrow}d^{\dagger}_{+\downarrow})
|0\rangle$, 
indicating the involved levels by the subscripts,
while the triplet states 
can be written as
$|t,1\rangle=d^{\dagger}_{+\uparrow}d^{\dagger}_{-\uparrow}|0\rangle$, 
$|t,0\rangle=\frac{1}{\sqrt{2}}(
d^{\dagger}_{+\uparrow}d^{\dagger}_{-\downarrow}+
d^{\dagger}_{-\uparrow}d^{\dagger}_{+\downarrow})
|0\rangle$,
$|t,-1\rangle=d^{\dagger}_{+\downarrow}d^{\dagger}_{-\downarrow}|0\rangle$, 
labeled 
by the $S_{z}$ quantum number and
the vacuum, $|0\rangle$, refers to the state where 
the investigated two levels are not occupied.

\begin{figure}
\includegraphics[width=170pt,clip=true]{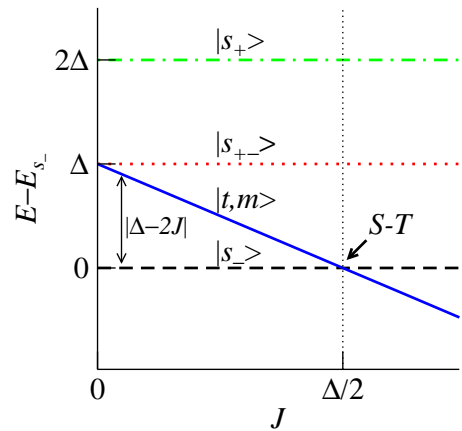}
\caption{The three singlet and three triplet states which can be constructed on the dot. The singlet states are labeled with the 
occupied levels, the triplet states with the $S_{z}$ value of the spin.}\label{statesj}
\end{figure}

The energies of these states can be distinguished by the values of the Hund's rule coupling ($J$) and the level-splitting ($\Delta$), 
while the on-site Coulomb $U$ gives equal contribution to all of the energies of these states. 
In Fig.~\ref{statesj} we can see the evolution of the energy-levels as a function of the Hund's rule coupling. 
Only the energy of the triplet states depend on $J$, because the energy term connected to Hund's rule coupling is $-JS(S+1)$,  
so it decreases the energy in the case of parallel spin alignment, $S=1$, and has no contribution when $S=0$. 
By the investigation of the spectral functions, we will change the value of $\Delta$ in the figures, while in the experiments the level splitting can be 
shifted by varying the parameters.\cite{Wiel,Kogan,granger} It can be seen in Fig.~\ref{statesj} that $\Delta/2=J$ is a special point considering 
the ground state. When $J<\Delta/2$, the ground state of the dot is non-degenerate singlet, but for $J>\Delta/2$ the 
threefold degenerate triplet gives the ground state. For $J=\Delta/2$ the $|s_{-}\rangle$ 
and the three triplet states ($|t,m\rangle$) are degenerate and therefore give a fourfold degenerate ground state. 
This is exactly true only for a bare dot, by the introduction of the hybridization, 
the transition between the ground states not occurs at the exact point of $\Delta/2=J$, but takes place in a wider range of the $\Delta$ 
parameter and becomes a crossover. The introduction of the hybridization leads the system to show Kondo-effect in the case of degenerate ground state. 

\begin{figure}
\includegraphics[width=240pt,clip=true]{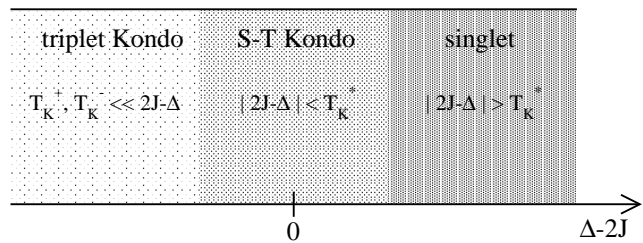}
\caption{Different regimes corresponding to different ground state configurations as a function of $\Delta$ for fixed $J$}\label{std}
\end{figure}

Fig.~\ref{std} shows what regimes can be observed in the quantum dot coupled to two conduction modes for different parameters.  
In this figure, the relevant parameter is $\Delta-2J$, while we take $J$ fixed and change $\Delta$. 
If we compare it with Fig.~\ref{statesj}
we can see that for a bare dot in the case of $0<\Delta-2J$ the ground state is a singlet, and the threefold degenerate triplet 
for $0>\Delta-2J$. Fig.~\ref{std} shows an intermediate regime also, where $|s_{-}\rangle$ and the three triplet states 
are degenerate, this would be the case only in the point $2J=\Delta$ for the bare dot, but for finite hybridization, this regime becomes finite.

The first regime of Fig.~\ref{std} with triplet ground state corresponds to triplet Kondo effect. 
In this case there is a finite $\vec{S}=1$ spin on the dot, but it is screened by the two conduction electron modes connected to the dot. 
Each mode screens $S=\frac{1}{2}$ from the whole spin, i. e. two separate Kondo effect screens the whole spin of the dot,
which can be characterized by 
different Kondo-temperatures, $T_{K}^{+}$ and $T_{K}^{-}$ because the hybridization-parameters, $\Gamma_{\pm}$, defined through 
Eq.~(\ref{gammadef}) can be different for the two levels. This phenomena is called two-stage Kondo effect.\cite{zarand,Wiel,granger} 

The second regime, connected to the singlet-triplet Kondo effect in Fig.~\ref{std} corresponds to the case where $|s_{-}\rangle$ 
and the three triplet states, $|t,m\rangle$ $m=-1,0,1$, gives an effectively degenerate ground state. 
They can be degenerate in a finite range of $\Delta$ because of the finite value of $\Gamma_{\pm}$. 
The Kondo-temperature corresponding to this situation, $T_{K}^{*}$, gives the energy scale of the singlet-triplet Kondo-effect. 
Generally $T_{K}^{*}$ is larger than 
$T_{K}^{\pm}$.\cite{eto} The system shows Kondo-effect for the parameters, 
$|2J-\Delta|<T_{K}^{*}$, which means considering the effect of the hybridization, that $2J$ is 
effectively equal to $\Delta$, the four above mentioned states can be considered degenerate. 

The third regime, called singlet, is corresponding to the case, where $|s_{-}\rangle$ is the ground state, when $|2J-\Delta|>T_{K}^{*}$ it 
cannot be effectively 
degenerate with any other states. The first excited state, which is the threefold degenerate triplet, 
can be reached by applying finite energy, this has a clear fingerprint in the spectral functions. 

\subsection{The symmetric case $(\Gamma_{+}=\Gamma_{-})$}\label{sym}

In this section we assume that the dot is completely symmetric, the diagonal elements of the $\ul{\ul{\Gamma}}$-matrix defined in 
Eq.~(\ref{gammadef}) are equal, which means that the tunneling matrix has the following structure,
\be
\ul{\ul{t}}=
\begin{bmatrix}
 t & t \\
 t & -t 
\end{bmatrix}\;.
\ee
We investigate the spectral functions for $J/U=0$ and $J/U=0.15$. In both of the cases the hybridization parameters are 
$\Gamma_{\pm}/U=0.785$. It is a clear feature of the FLEX spectral functions, that the FLEX approximation cannot give back the 
charge excitation peaks,\cite{white} similarly to all of the approximations which calculate the end Green's function by iterating 
the fully interacting 
Green's function in every step.\cite{noneqand1,noneqand2,noneqand3,noneqand4,white} 
Otherwise such treatments are the only candidates to be \textit{conserving}, we assume that the above mentioned 
feature doesn't influence the description of the transport properties too much. 
Beside this feature, the FLEX approximation describes well the width of 
the Kondo-peak, which is connected to the Kondo energy scale and what is assumed to be relevant by the description of the transport properties. 

\begin{figure}
\includegraphics[width=240pt,clip=true]{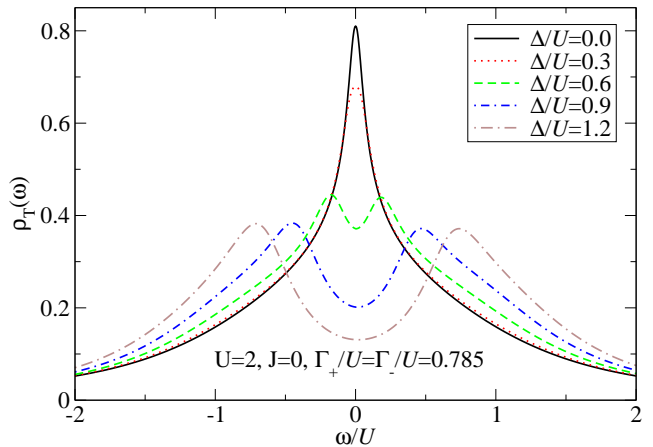}
\caption{Total spectral function, $\rho_{T}(\omega)$ for $J/U=0$ and $\Gamma_{\pm}/U=0.785$ for different values of level splitting, 
$\Delta/U$}\label{spect0}
\end{figure}

Considering first the spectral functions of Fig.~\ref{spect0} for $J/U=0$ and different $\Delta/U$ parameters. 
We have to note at the beginning that in the lack of finite $J$, 
only two regimes can be investigated from the three regimes of Fig.~\ref{std}, 
as we change $\Delta$, the triplet Kondo effect is not present for $J/U=0$. 
For $\Delta/U=0$ and small level splitting values ($\Delta<T_{K}^{*}$), 
the spectral functions show the existence of a singlet-triplet Kondo effect, 
while they show a one-peak structure with a peak at the Fermi-energy, which is the Kondo-peak. 
Above some value of $\Delta/U$, the Kondo-peak splits. 
For large $\Delta/U$ values the two side-peaks correspond to the triplet excitations, and they are residing on the  
exact place of $\omega=\pm\Delta/2$. With this finite amount of energy we can excite the first excited state which is the threefold 
degenerate triplet in this case. For intermediate $\Delta/U$ values ($\Delta/U=0.6$ in Fig.~\ref{spect0}) the splitting is not corresponding 
unambiguously to the triplet excitations, the peak positions are influenced by the Kondo effect also, while the system is in the transition regime 
between the singlet-triplet Kondo and singlet regimes. 

\begin{figure}
\includegraphics[width=240pt,clip=true]{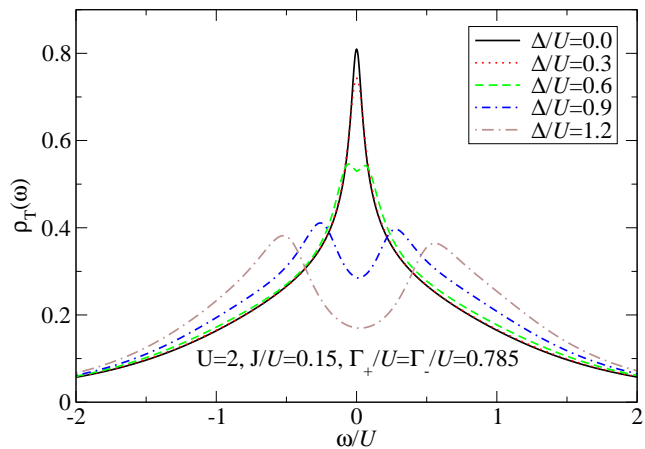}
\caption{Total spectral function, $\rho_{T}(\omega)$ for $J/U=0.15$ and $\Gamma_{\pm}/U=0.785$ for different values of level splitting, 
$\Delta/U$}\label{spectj}
\end{figure}

In the presence of finite Hund's rule coupling, $J$, the spectral function is shown on Fig.~\ref{spectj} for $J/U=0.15$. 
In this set of spectral functions we can see differences compared to Fig.~\ref{spect0}. By the introduction a finite value of $J$, we can 
investigate the triplet Kondo-effect for $T_{K}^{\pm}<<2J-\Delta$ ($T_{K}^{+}=T_{K}^{-}$ for $\Gamma_{+}=\Gamma_{-}$), 
which is true for small values of $\Delta$ for the spectral functions of Fig.~\ref{spectj}. 
Which was assumed earlier, that $T_{K}^{\pm}$ is smaller than $T_{K}^{*}$, can be clearly seen for $\Delta/U=0$ 
in Fig.~\ref{spectj}, i.e. the width of the triplet Kondo-peak is smaller than the width of the singlet-triplet Kondo peak (the effect is larger for 
larger values of $J$). 
For large values of $\Delta$ (e.g. $\Delta/U=1.2$ in Fig.~\ref{spectj}), we can see the above detailed fingerprints of singlet ground state, the 
two triplet side peaks at $\omega/U=\pm(\Delta-2J)/U$. For intermediate $\Delta$ values the spectral function is influenced by 
the singlet-triplet Kondo-effect, when $|2J-\Delta|<T_{K}^{*}$. For $\Delta/U=0.6$ this is the case, and the corresponding spectral function on 
Fig.~\ref{spectj} has a wider Kondo-peak than for $\Delta/U=0$, as it was assumed.

\subsection{The asymmetric case, $\Gamma_{+}\neq\Gamma_{-}$}\label{assym}

\begin{figure}
\includegraphics[width=240pt,clip=true]{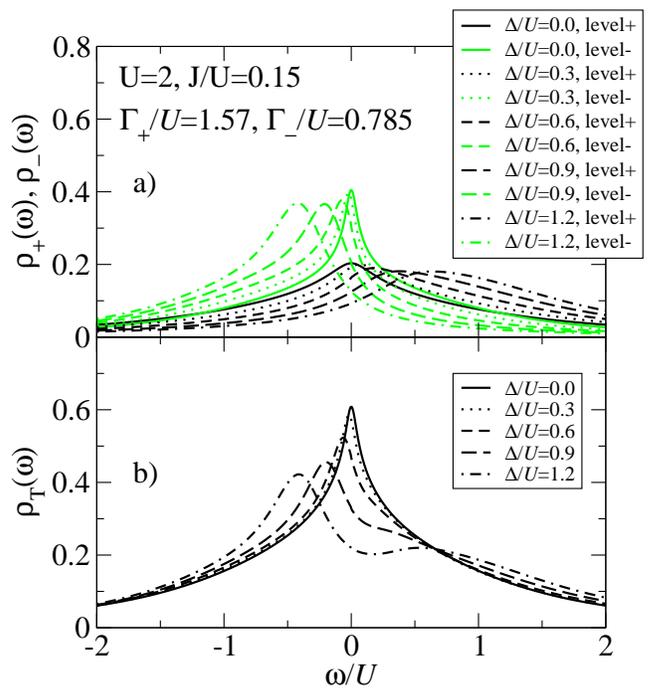}
\caption{The spectral functions for the levels separately ($a$), and the total spectral function, $\rho_{T}(\omega)$ for $J/U=0.15$, 
$\Gamma_{+}/U=1.57$ and $\Gamma_{-}/U=0.785$ for different 
values of level splitting, $\Delta/U$}\label{spect_as}
\end{figure}

In the case $\Gamma_{+}\neq\Gamma_{-}$, there is no particle-hole symmetry in the system. The value of the total occupation of the quantum dot is 
not exactly two any more, and the spectral function is not symmetric. 
For $\Gamma_{+}\neq\Gamma_{-}$ we consider only the physically interesting case, when the Hund's rule coupling is finite. 
The spectral functions separately for the levels for parameters $\Gamma_{+}/U=1.57$, $\Gamma_{-}/U=0.785$ and $J/U=0.15$ can be 
seen in Fig.~\ref{spect_as}a  
while in Fig.~\ref{spect_as}b the total spectral functions is observable. 
For $\Delta/U=0$ similarly to the $\Gamma_{+}=\Gamma_{-}$ case of Fig.~\ref{spectj}, we can see the fingerprints of a triplet Kondo effect. 
If we consider the $\Delta/U=0$ spectral functions for each level in Fig.~\ref{spect_as}a, we can see the two different width of the 
Kondo-peaks due to the different Kondo-temperatures. This feature shows the dependence of $T_{K}^{\pm}$ on the hybridization parameters, $\Gamma_{\pm}$. 
For small splitting, this is the cause of the asymmetry of the total spectral function, but while the triplet excitation peaks also have different 
width, the asymmetry is characteristic for all of $\Delta/U$ values. 
The trends with increasing $\Delta$ can be clearly observed on the separately shown spectral 
functions of Fig.~\ref{spect_as}a. It is transparent in that figure, that when the Kondo-peak changes little and it leaves the Fermi-energy 
slightly then the system shows triplet or singlet-triplet Kondo-effect. At the $\Delta$ value where the Kondo-peak gets wider, that is the 
point of transition from triplet to singlet-triplet Kondo effect. For $\Delta/U=1.2$ where the peak of the spectral functions in 
Fig.~\ref{spect_as}b far from $\omega=0$ and from the slightly splitted peaks, and have larger width than the peaks corresponding to Kondo effect, 
we can assume, that these peaks corresponds to the triplet excitations, and the ground 
state of the quantum dot is a singlet for this parameter set.

\section{Conclusions}\label{conclu}

In this paper we applied fluctuation-exchange approximation to the non-equilibrium Anderson model. We generalized the FLEX approximation 
using Keldysh theory. The spectral functions were investigated for different parameter sets, we could catch singlet-triplet transition 
by varying $\Delta$ parameter, and could observe triplet and singlet-triplet Kondo effect on the spectral functions. 
It is relevant to note as an evidence of the good description of the Kondo energy scales, that the width of the spectral functions corresponding to 
triplet and singlet-triplet Kondo effect are distinguishable. The good description of the width of the peaks is also relevant by distinguishing 
the triplet excitation peaks from the peaks corresponding to singlet-triplet Kondo effect. 
We can conclude that although the FLEX approximation does not give back the so-called charge side-peaks in the spectral functions, 
the description of the behavior the central Kondo-peak is satisfactory for all parameters. 

\section{Acknowledgement}

This research has been supported by Hungarian Grant, OTKA NN76727. 

\appendix
\section{Calculation of the hybridized Green's function}\label{hybrg}

We can construct the hybridized Green's functions in some easy steps. 
The inverse of the atomic Green's function is (independent from spin)
\be
 {{g^{(0)}}^{-1}}_{i\sigma\kappa}^{j\sigma\kappa'}(\omega) = \left(\omega\delta_{ij}-\varepsilon_{ij}\right)\delta_{\kappa\kappa'}S(\kappa)\;,
\ee
where $S(\kappa)$ is the Keldysh sign. The hybridization self-energy is (also independent from spin)
\be
 \Sigma_{i\sigma\kappa}^{j\sigma\kappa'} = \sum_{\alpha\in L,R}t_{\alpha i}t_{\alpha j}\Sigma_{\alpha}^{\kappa\kappa'}\;,
\ee
where
\bea
 \Sigma_{\alpha}^{++}(\omega) & = & \pi i(2f_{\alpha}(\omega)-1)\;,\\
 \Sigma_{\alpha}^{+-}(\omega) & = & -2\pi if_{\alpha}(\omega)\;,\\
 \Sigma_{\alpha}^{-+}(\omega) & = & -2\pi i(f_{\alpha}(\omega)-1)\;,\\
 \Sigma_{\alpha}^{--}(\omega) & = & \pi i(2f_{\alpha}(\omega)-1)\;,
\eea
where $f_{\alpha}(\omega)=f(\omega-\mu_{\alpha})$ is the Fermi function. The hybridized Green's function is
\be
 {g^{-1}}_{i\sigma\kappa}^{j\sigma\kappa'} = \left(\omega\delta_{ij}-\varepsilon_{ij}\right)\delta_{\kappa\kappa'}S(\kappa) - 
\sum_{\alpha\in L,R}t_{\alpha i}t_{\alpha j}\Sigma_{\alpha}^{\kappa\kappa'}\;,
\ee
independent and therefore diagonal in '$\sigma$'. The indices '$i\sigma\kappa$' will be taken to a compound index '$\alpha$'.

\section{Simplification of the self-energy calculation}\label{selfapp}

Here we show how can we use the simple structure of the polarization and interaction vertex matrix to somehow simplify the self-energy 
expressions. We can use a fact that multiplication of $\ul{\ul{\Pi}}_{s,t}$ and $\ul{\ul{\tilde{\Gamma}}}_{s,t}$ will also have 
singlet or triplet structure. 
The simplifications shown here can be used by each term of self energy which contain polarization and$/$or vertex matrices, we 
will show the simplification step in the FLEX self energy, 
\be
 {\Sigma_{FLEX}}_{\alpha}^{\beta}(t) = -\sum_{\tilde{\alpha}\tilde{\beta}}
\left(\ul{\ul{\tilde{\Gamma}}}\ul{\ul{\Pi}}(t)\ul{\ul{\tilde{\Gamma}}}\right)_{\alpha\tilde{\alpha}}^{\beta\tilde{\beta}}
G_{\tilde{\alpha}}^{\tilde{\beta}}(t)\;.\label{appflex}
\ee
Because the self-energy is independent of the spin, we are free to fix spin e. g. in $\uparrow$. Only considering the spin arrangements in 
the above expression
\be
\Sigma_{\uparrow}^{\uparrow} = -\left(\ul{\ul{\tilde{\Gamma}}}\ul{\ul{\Pi}}(t)\ul{\ul{\tilde{\Gamma}}}\right)
_{\uparrow\uparrow}^{\uparrow\uparrow}
G_{\uparrow}^{\uparrow}+
(-i)^{2}\left(\ul{\ul{\tilde{\Gamma}}}\ul{\ul{\Pi}}(t)\ul{\ul{\tilde{\Gamma}}}\right)
_{\uparrow\downarrow}^{\uparrow\downarrow}G_{\downarrow}^{\downarrow}\;,
\ee
the multiplications in the brackets can be expressed by the singlet and triplet terms of the multiplication,
\bea
\left(\ul{\ul{\tilde{\Gamma}}}\ul{\ul{\Pi}}(t)\ul{\ul{\tilde{\Gamma}}}\right)_{\uparrow\uparrow}^{\uparrow\uparrow} & = & 
\frac{1}{2}\left(\left(\ul{\ul{\tilde{\Gamma}}}\ul{\ul{\Pi}}(t)\ul{\ul{\tilde{\Gamma}}}\right)_{s}+
\left(\ul{\ul{\tilde{\Gamma}}}\ul{\ul{\Pi}}(t)\ul{\ul{\tilde{\Gamma}}}\right)_{t}\right)\;,\\ 
 \left(\ul{\ul{\tilde{\Gamma}}}\ul{\ul{\Pi}}(t)\ul{\ul{\tilde{\Gamma}}}\right)_{\uparrow\downarrow}^{\uparrow\downarrow} & = & 
\left(\ul{\ul{\tilde{\Gamma}}}\ul{\ul{\Pi}}(t)\ul{\ul{\tilde{\Gamma}}}\right)_{t}\;.
\eea
The Green's functions are not spin dependent so we can simplify the original expression of the self energy (\ref{appflex}) by a form which 
uses the easily computable singlet and triplet contributions of the polarization and vertex matrices, 
\bea
& & {\Sigma_{FLEX}}_{\alpha}^{\beta}(t) = \nonumber\\
& = & -\sum_{\tilde{\alpha}\tilde{\beta}}
\left(\frac{3}{2}\left(\ul{\ul{\tilde{\Gamma}}}\ul{\ul{\Pi}}(t)\ul{\ul{\tilde{\Gamma}}}\right)_{t}+
\frac{1}{2}\left(\ul{\ul{\tilde{\Gamma}}}\ul{\ul{\Pi}}(t)\ul{\ul{\tilde{\Gamma}}}\right)_{s}\right)_{\alpha\tilde{\alpha}}^{\beta\tilde{\beta}}
G_{\tilde{\alpha}}^{\tilde{\beta}}(t)\;.
\eea
This means that we have to work only 
with the level and Keldysh indices by the calculation of the self-energy, which acts as a very large simplification by the storage of the 
matrices.


\begin{thebibliography}{}
\bibitem{us} B. Horv\'ath, B. Lazarovits and G. Zar\'and (preprint on cond-mat)
\bibitem{rev1} A. C. Hewson, \textit{The Kondo Problem to Heavy Fermions} (Cambridge University Press, Cambridge, 1993).
\bibitem{rev2} L. P. Kouwenhoven, C. M. Marcus, P. L. McEuen, S. Tarucha, R. M. Westervelt
and N. S .Wingreen, Electron transport in quantum dots Mesoscopic Electron Transport ed L. L. Sohn,
G. Sch\"on and L. P. Kouwenhoven (Kluwer Series E vol 345) (1996). 
\bibitem{rev3} L. P. Kouwenhoven, D. G. Austing and S. Tarucha, Rep. Prog. Phys. {\bf 67}, 701 (2001).
\bibitem{rev4} L. I. Glazman and M. Pustilnik, "Low temperature transport through a quantum dot" in 
"Nanophysics: Coherence and Transport," eds. H. Bouchiat et al. (Elsevier, 2005)
\bibitem{rev5} G. Zar\'and, Phil. Mag. {\bf 86}, 2043 (2006).
\bibitem{rev6} M. Grobis, I.G. Rau, R.M. Potok, and D. Goldhaber-Gordon. "Kondo Effect in Mesoscopic Quantum Dots", 
Handbook of Magnetism and Magnetic Materials, H. Kronmüller and S. Parkin, eds., (Wiley, 2007).
\bibitem{noneqand0} B. Horv\'ath, B. Lazarovits, O. Sauret, and G. Zar\'and, Phys. Rev. B {\bf 77}, 113108 (2008). 
\bibitem{noneqand1} K. S. Thygesen and A. Rubio, J. Chem. Phys. {\bf 126}, 091101 (2007). 
\bibitem{noneqand2} K. S. Thygesen and A. Rubio, Phys. Rev. B {\bf 77}, 115333 (2008). 
\bibitem{noneqand3} X. Wang, C. D. Spataru, M. S. Hybertsen and A. J.  Millis, Phys. Rev. B {\bf 77}, 045119 (2008). 
\bibitem{noneqand4} C. D. Spataru, M. S. Hybertsen, S. G. Louie and A. J. Millis, Phys. Rev. B {\bf 79}, 155110 (2009). 
\bibitem{noneqand5} S.-P. Chao and G. Palacios arXiv:cond-mat/1003.5395 (2010). 
\bibitem{noneqand6} P. R. Bas and A. A. Aligia, J. Phys.:Condens. Matter {\bf 22}, 025602 (2010). 
\bibitem{noneqand7} D. E. Logan, Ch. J. Wright, and M. R. Galpin, Phys. Rev. B {\bf 80}, 125117 (2009). 
\bibitem{noneqand8} A. L. Yeyati, A. Mart\'in-Rodero, and F. Flores, Phys. Rev. Lett. {\bf 71}, 2991 (1993). 
\bibitem{aligia} A. A. Aligia, Phys. Rev. B {\bf 74}, 155125 (2006).
\bibitem{Wiel} W. G. van der Wiel, S. De Franceschi, J. M. Elzerman, S. Tarucha, L. P. Kouwenhoven, J. Motohisa, F. Nakajima, and T. Fukui, 
Phys. Rev. Lett. {\bf 88}, 126803. (2002).
\bibitem{Kogan} A. Kogan, G. Granger, M. A. Kastner, D. Goldhaber-Gordon and H. Shtrikman, Phys. Rev. B {\bf 67} 113309 (2003).
\bibitem{granger} G. Granger, M. A. Kastner, Iuliana Radu, M. P. Hanson and A. C. Gossard, Phys. Rev. B {\bf 72}, 165309 (2005).
\bibitem{kotliar} H. Kajueter and G. Kotliar, Phys. Rev. Lett. {\bf 77}, 131 (1996)
\bibitem{sasaki} S. Sasaki, S. De Franceschi, J. M. Elzerman, W. G. van der Wiel, M. Eto, S. Tarucha and L. P. Kouwenhoven, Nature {\bf 405}, 764 (2000).
\bibitem{zarand} W. Hofstetter and G. Zar\'and, Phys. Rev. B {\bf 69}, 235301 (2004).
\bibitem{roch} N. Roch, S. Florens, V. Bouchiat, W. Wernsdorfer and F. Balestro, Nature {\bf 453}, 633 (2008).
\bibitem{yeyati2} A. L. Yeyati, F. Flores and A. Martin-Rodero, Phys. Rev. Lett. {\bf 83} 600 (1999).
\bibitem{rammer} J. Rammer and H. Smith, Rev. Mod. Phys. {\bf 58}, 323 (1986).
\bibitem{andrei} P. Mehta and N. Andrei, Phys. Rev. Lett {\bf 96}, 216802 (2006). 
\bibitem{anders} F. B. Anders and A. Schiller, Phys. Rev. Lett. {\bf 95}, 196801 (2005).
\bibitem{egger} J. Eckel, F. Heidrich-Meisner, S. G. Jakobs, M. Thorwart, M. Pletyukhov and R. Egger, New. J. Phys. {\bf 12}, 043042 (2010).
\bibitem{Scholl} A. Holzner, I. P. McCulloch, U. Schollwöck, J. von Delft and F. Heidrich-Meisner, Phys. Rev. B {\bf 80}, 205114 (2009). 
\bibitem{freericks} J. K. Freericks, V. M. Turkowski and V. Zlatic, Phys. Rev. Lett. {\bf 97}, 266408 (2006).
\bibitem{bickers} N. E. Bickers, D. J. Scalapino and S. R. White, Phys. Rev. Letters {\bf 62}, 961 (1989); 
N. E. Bickers and D. J. Scalapino, Ann. Phys. (N.Y.) {\bf 193}, 206 (1989); 
N. E. Bickers and S. R. White, Phys. Rev. B {\bf 43}, 8044 (1991). 
\bibitem{kadanoff} G. Baym and L. P. Kadanoff, Phys. Rev. {\bf 124}, 287 (1961). 
\bibitem{baym} G. Baym, Phys. Rev. {\bf 127}, 1391 (1962). 
\bibitem{white} J. A. White, Phys. Rev. B {\bf 45}, 1100 (1992).
\bibitem{Pust1} M. Pustilnik and L. I. Glazman, Phys. Rev. Lett. {\bf 87}, 216601 (2001).
\bibitem{Pust2} M. Pustilnik and L. I. Glazman, Phys. Rev.B {\bf 64}, 045328 (2001). 
\bibitem{izumida} W. Izumida, O. Sakai and Y. Shimizu, J. Phys. Soc. Jpn. {\bf 67}, 2444 (1998).
\bibitem{eto} M. Eto and Y. V. Nazarov, Phys. Rev. Lett. {\bf 85}, 1306 (2000); 
M. Eto, J. Phys. Soc. Jpn. {\bf 74}, 95 (2005).




\bibliography{apssamp}
\end{thebibliography}
\end{document}